%% file: Fren_v12.tex
\documentclass[twocolumn,11pt]{article}
\usepackage{times}
\input prooftree
\newtheorem{definition}{Definition}

\begin{document}
\global\def\refname{{\normalsize \it References:}}
\baselineskip 12.5pt
%
%
%
\title{\LARGE \bf ImNet: An Imperative Network Programming Language}

\date{}

\author{\hspace*{-10pt}
\begin{minipage}[t]{3.4in} \normalsize \baselineskip 12.5pt
\centerline{Mohamed A. El-Zawawy$^{1,2,}$\footnote{Corresponding
author.}}\centerline{$^1$College of Computer and Information
Sciences} \centerline{Al Imam Mohammad Ibn Saud Islamic University}
\centerline{ (IMSIU)} \centerline{Riyadh }\centerline{ Kingdom of
Saudi Arabia}\centerline{} \centerline{$^2$Department of
Mathematics} \centerline{Faculty of Science} \centerline{Cairo
University} \centerline{Giza 12613}
\centerline{Egypt}\centerline{maelzawawy@cu.edu.eg}
\end{minipage} \kern 0in
\begin{minipage}[t]{3.3in} \normalsize \baselineskip 12.5pt
\centerline{Adel I. AlSalem} \centerline{College of Computer and
Information Sciences} \centerline{Al Imam Mohammad Ibn Saud Islamic
University} \centerline{ (IMSIU)}  \centerline{Riyadh }\centerline{
Kingdom of Saudi Arabia} \centerline{alsalem@ccis.imamu.edu.sa}
\end{minipage}
%
%
\\ \\ \hspace*{-10pt}
\begin{minipage}[b]{6.9in} \normalsize
\baselineskip 12.5pt {\it Abstract:} One of the most recent
architectures of networks is Software-Defined Networks (SDNs) using
a controller appliance to control the set of switches on the
network. The controlling process includes installing or uninstalling
packet-processing rules on flow tables of switches.\\
This paper presents a high-level imperative network programming
language, called \textit{ImNet}, to facilitate writing efficient,
yet simple, programs executed by controller to manage switches.
\textit{ImNet} is simply-structured, expressive, compositional, and
imperative. This paper also introduces an operational semantics to
\textit{ImNet}. Detailed examples of programs (with their
operational semantics) constructed in \textit{ImNet} are illustrated
in the paper as well.
\\ [4mm] {\it Key--Words:}
Network programming languages, controller-switch architecture,
operational semantics, syntax, \textit{ImNet}.
\end{minipage}
\vspace{-10pt}}

\maketitle

\thispagestyle{empty} \pagestyle{empty}
%
%

\section{Introduction}
\label{intro}\vspace{-4pt}

A network is a group of appliances connected to exchange data. Among
these appliances are switches forwarding data depending on MAC
addresses, routers forwarding data depending on IP addresses, and
firewalls taking care of forbidden data. The network appliances are
connected using a model that efficiently allows  forwarding,
storing, ignoring, tagging, and providing statistics about data
moving in the network. Some of the network appliances, like
routers~\cite{Arneson,Suzuki}, are special in their functionality as
they have some control over the network. This enables routers to
compute and determine routes of data in the network. Of course
different networks have different characteristics and abilities.

In 2011, the Open Networking Foundation~\cite{website}, suggested
removing the control owned by different network appliances and
adding, instead, a general-purpose appliance, controller, to program
different network appliances and querying data flowing in the
network. The impact of this simple suggestion is huge; giant
networks do not need special-purpose, complex, expensive switches
any more. In such networks, cheap programmable switches can be used
and programmed to configure and optimize networks via writing
programs~\cite{Rexford} running on controllers.

Software-Defined Networks (SDNs)~\cite{Foster13} are networks
established using the controller-switch architecture. A precise
implementation of this architecture is OpenFlow~\cite{Cai10} used to
achieve various network-wide applications such as monitoring data
flow, balancing switch load, network management, controlling
appliances access, detection of service absence, host mobility, and
forwarding data center. Therefore SDNs caused the appearance of
network programming languages~\cite{Monsanto,Hong,Elsts,Bain}.

This paper presents \textit{ImNet}, an imperative high-level network
programming language. \textit{ImNet}  expresses commands enabling
controllers to program other network appliances including switches.
\textit{ImNet} has a clear and simply-structured syntax based on
classical concepts of imperative programming that allows building
rich and robust network application in a natural way. \textit{ImNet}
can be realized as a generalization of Frenetic~\cite{Foster10}
which is a functional network programming language. This is clear by
the fact that the core of programs written in \textit{ImNet} and
Frenetic is based on a query result in the form of stream of values
(packets, switches IDs, etc.). Commands for treating packets in
\textit{ImNet} include constructing and installing (adding to flow
tables of switches) switch rules. \textit{ImNet} supports building
simple programs to express complex dynamic functionalities like load
balancing and authentication.  \textit{ImNet} programs can also
analyze packets and historical traffic patterns.

\begin{figure*}[t]
\centering \fbox{
\begin{minipage}{11.5 cm}
{\footnotesize{
\begin{eqnarray*}
& &x\in \hbox{lVar}\qquad\qquad Q\in \hbox{Queries}\\
{et} \in \hbox{Eventrans}   &::= & {\hbox{Lift} (x,\lambda
t.f(t))}\mid {\hbox{ApplyLft} (x,\lambda t.f(t))}\mid
{\hbox{ApplyRit}
(x,\lambda t.f(t))}\mid \\
&& \hbox{Merge}(x_1,x_2)\mid
\hbox{MixFst}(A,x_2,x_3)\mid\hbox{MixSnd}(A,x_2,x_3)\mid \\ & &
\hbox{Filter}(x,\lambda.f(t))\mid \hbox{Once}(x)
\mid \hbox{MakForwRule}(x)\mid\hbox{MakeRule}(x)\\
S \in \hbox{Stmts}   &::= &  {x:= {et}}\mid S_1; S_2\mid
\hbox{AddRules}(x)\mid \hbox{Register}\mid \hbox{Send}(x)
\\
D\in \hbox{Defs}   &::= & \epsilon\mid {x:= Q}\mid D D.
\\
p\in \hbox{Progs}   &::= & D\gg S.
\end{eqnarray*}
}}\caption{ImNet Syntax.}\label{f1}
\end{minipage}
}
\end{figure*}

\subsection*{Motivation}
\vspace{-4pt} The motivation of this paper is the lack of a simple
syntax for an imperative network programming language. Yet, a
stronger motivation is that most existing network programming
languages are not supported theoretically (using operational
semantics, type systems, program logics like Floyd–Hoare
logic,etc.).

\subsection*{Contributions}
\vspace{-4pt}

Contributions of this paper are the following.
\begin{enumerate}
\item  A new simply-structured syntax for an imperative network
programming language; \textit{ImNet}.
\item An operational semantics (in the form of states and inference rules)
for constructs of \textit{ImNet}.
\item Two detailed examples of programs constructed in \textit{ImNet}
with their precise operation semantics.
\end{enumerate}

\subsubsection*{Organization}
\vspace{-4pt}

The rest of this paper is organized as following. Section~\ref{s1}
presents the syntax and semantics of \textit{ImNet}. The proposed
semantics is operational and hence consists of states and inference
rules presented in Section~\ref{s1}. Two detailed examples of
programmes built in \textit{ImNet} are presented in
Section~\ref{s2}. This section also explains how the two examples
can be assigned precise semantics using our proposed operational
semantics. Section~\ref{s3} reviews related work and gives
directions for future work. Section~\ref{s4} concludes the paper.

\section{Semantics}
\label{s1}\vspace{-4pt}

This section presents the syntax and semantics of \textit{ImNet}, a
high-level programming language for SDN networks using
switch-controller architecture. Figure~\ref{f1} shows the syntax of
\textit{ImNet}. Figures~\ref{f2} and~\ref{f3} present the semantics
of \textit{ImNet} constructs. The proposed semantics is operational
and its states are defined in the following definition.

\begin{definition}\label{state}
\begin{enumerate}
\item $t\in$ Types = \{int, Switch IDs, Packet,
(Switch IDs, int, bool)\}$\cup\{(t_1,t_2)\mid t_1,t_2\in
\hbox{Types}\}$.
\item $v\in\hbox{Values} =$ Natural numbers $\ \cup \hbox{ Switch IDs }
 \cup\hbox{Packets}\cup \hbox{Switch IDs }\times $ Natural numbers $\times $
 Boolean values\ $\cup\
 \{(v_1,v_2)\mid v_1,v_2\in \hbox{Values}\}$.
The expression $v:t$ denotes that the type of the value $v$ is $t$.
\item ${ev}\in \hbox{Events} = \{(v_1,v_2,\ldots,v_n)\mid
\exists t (\forall i\ v_i:t)\}$.
\item Actions= {\{sendcontroller, sendall, sendout, change(h,v) \}}.
\item $r\in\hbox{Rules} =\hbox{Patterns} \times \hbox{Actions}$.
\item ${rl}\in \hbox{Rule-lists}=\{[r_1,r_2,\ldots,r_n]\mid r_i\in\hbox{Rules}\}$.
\item ${ir}\in \hbox{Intial-rule-assignment}=\hbox{Switch IDs}\times \hbox{Rules}$.
\item $\sigma\in\hbox{Swich-states}=\hbox{Flow-tables} =\hbox{Switch IDs}
\rightarrow \hbox{Rule-lists}$.
\item $\gamma\in\hbox{Variable-states} =\hbox{Var} \rightarrow
\hbox{Events}\cup \hbox{Rule-lists} $.
\item $s\in \hbox{States} =\hbox{Swich-states}\times
\hbox{Variable-states}\times\hbox{Rule-lists}. $
\end{enumerate}
\end{definition}

A program in \textit{ImNet} is a sequence of queries followed by a
statement. The result of each query is an event which is a finite
sequence of values. The event concept is also used in Frenetic.
However an event in Frenetic is an infinite sequence of values. A
value is an integer, a switch ID, a packet, a triple of a switch ID,
an integer, and a Boolean value, or a pair of two values. Each value
has a type of the set \textit{Types}. In this paper, we focus on the
details of statements as this is the most interesting part in a
network programming language.

Possible actions taken by a certain switch on a certain packet are
\textit{sendcontroller}, \textit{sendall}, \textit{sendout}, or
\textit{change(h,v)}. The action \textit{sendcontroller} sends a
packet to the controller to take care of it. The action
\textit{sendall} sends the packet to all other switches. The action
\textit{sendout} sends the packet out of the switch through a
certain port. The action \textit{change(h,v)} modifies the header
field $h$ of the packet to the new value $v$.

\begin{figure*}[t]
\centering \fbox{
\begin{minipage}{15cm}
{\footnotesize{
\[
\begin{prooftree}
v_i:t \qquad \gamma(x)=(v_1,v_2,\ldots,v_n)\justifies \hbox{Lift}
(x,\lambda t.f(t)):\gamma\rightarrow
(f(v_1),f(v_2),\ldots,f(v_n))\thickness=0.08em\using{(\hbox{Lift}^s)}
\end{prooftree}
\]
\[
\begin{prooftree}
\gamma(x_1)=(v_1,v_2,\ldots,v_n)\qquad
\gamma(x_2)=(w_1,w_2,\ldots,w_n)\justifies
\hbox{Merge}(x_1,x_2):\gamma\rightarrow
((v_1,w_1),(v_2,w_2),\ldots,(v_n,w_n))\thickness=0.08em\using{(\hbox{Merge}^s)}
\end{prooftree}
\]
\[
\begin{prooftree}
\gamma(x)=(v_1,v_2,\ldots,v_n)\qquad A=\{i\mid f(v_i)=\hbox{true}\}
\justifies \hbox{Filter}(x,\lambda.f(t)):\gamma\rightarrow
(\dots,v_i,\ldots\mid i\in A)
\thickness=0.08em\using{(\hbox{Filter}^s)}
\end{prooftree}
\]
\[
\begin{prooftree}
v_i:t\qquad
\gamma(x)=((v_1,v_1^\prime),(v_2,v_2^\prime),\ldots,(v_n,v_n^\prime))\justifies
\hbox{ApplyLft} (x,\lambda t.f(t)):\gamma\rightarrow
((f(v_1),v_1^\prime),(f(v_2),v_2^\prime),\ldots,(f(v_n),v_n^\prime))\thickness=0.08em
\using{(\hbox{App}_1^s)}
\end{prooftree}
\]
\[
\begin{prooftree}
v_i^\prime:t\qquad
\gamma(x)=((v_1,v_1^\prime),(v_2,v_2^\prime),\ldots,(v_n,v_n^\prime))\justifies
\hbox{ApplyRit} (x,\lambda t.f(t)):\gamma\rightarrow
((v_1,f(v_1^\prime)),(v_2,f(v_2^\prime)),\ldots,(v_n,f(v_n^\prime)))\thickness=0.08em\using{(\hbox{App}_2^s)}
\end{prooftree}
\]
\[
\begin{prooftree}
\hbox{type}(x)\in \hbox{Types}
\justifies\hbox{Once}(x):\gamma\rightarrow
(\underbrace{x,x,\ldots,x}_{n\_times})
\thickness=0.08em\using{(\hbox{Once}^s)}
\end{prooftree}
\]
\[
\begin{prooftree}
\gamma(x_1)=(v^1_1,v^1_2,\ldots,v^1_n)\quad
\gamma(x_2)=(v^2_1,v^2_2,\ldots,v^2_n)\quad
A_1=A\cup\{v^1_1\}\quad\forall i>1. A_i=A_{i-1}\cup \{v^1_i\}
\justifies \hbox{MixFst}(A,x_1,x_2):\gamma\rightarrow
((A_1,v^2_1),(A_2,v^2_2),\ldots,(A_n,v^2_n))
\thickness=0.08em\using{(\hbox{Mix}_1^s)}
\end{prooftree}
\]
\[
\begin{prooftree}
\gamma(x_1)=(v^1_1,v^1_2,\ldots,v^1_n)\quad
\gamma(x_2)=(v^2_1,v^2_2,\ldots,v^2_n)\quad
A_1=A\cup\{v^2_1\}\quad\forall i>1. A_i=A_{i-1}\cup \{v^2_i\}
\justifies \hbox{MixSnd}(A,x_1,x_2):\gamma\rightarrow
((v^1_1,A_1),(v^1_2,A_2),\ldots,(v^1_n,A_n))
\thickness=0.08em\using{(\hbox{Mix}_2^s)}
\end{prooftree}
\]
\[
\begin{prooftree}
\gamma(x)=((v^1_1,a_1,v^3_1),(v^1_2,a_2,v^2_2),\ldots,(v^1_n,a_n,v^2_n))
\justifies \hbox{MakForwRule}(x):\gamma\rightarrow
[(v^1_1,(v^3_1,sendout(v^2_1))),(v^1_2,(v^3_2,sendout(v^2_2))),
\dots,(v^1_n,(v^3_n,sendout(v^2_n)))]
\thickness=0.08em\using{(\hbox{MFR}^s)}
\end{prooftree}
\]
\[
\begin{prooftree}
\gamma(x)=((v^1_1,a_1,v^2_1),(v^1_2,a_2,v^2_2),\ldots,(v^1_n,a_n,v^2_n))
\justifies \hbox{MakeRule}(x):\gamma\rightarrow
[(v^1_1,a_1(v^1_2)),(v^2_1,a_2(v^2_2)),\dots,(v^i_n,a_n(v^n_2))]
\thickness=0.08em\using{(\hbox{MkRl}^s)}
\end{prooftree}
\]
}} \caption{Operational semantics for event functions of
ImNet}\label{f2}
\end{minipage}
}
\end{figure*}

A rule in our semantics is a pair of \textit{pattern} and
\textit{action} where \textit{pattern} is a form that concretely
describes a set of packets and \textit{action} is the action to be
taken on elements of this set of packets. Rules are stored in tables
(called \textit{flow tables}) of switches.
\textit{Intial-rule-assignment} represents an initial assignment of
rules to flow tables of switches.

A state in the proposed operational semantics is a triple
$(\sigma,\gamma,ir)$. In this triple $\gamma$ captures the current
state of the program variables and hence is a map from the set of
variables to the set of events and rule lists. This is so because in
\textit{ImNet} variables may contain events or rule lists. The
symbol $\sigma$ captures the current state of flow tables of
switches and hence is a map from switches \textit{IDs} to rule
lists. Finally, $ir$ is an initial assignment of rules assigned to
switches but have not been registered yet (have not been added to
$\gamma$ yet).

There are five type of statements in \textit{ImNet}. The assignment
statement $x:= {ef}$ assigns the result of an event transformer (et)
to the variable $x$. The statement \textit{AddRules(x)} adds the
switch rules stored in $x$ to the reservoir of initially assigned
rules. These are rules that are assigned to switches but are not
added to flow tables yet. The statement \textit{Register} makes the
initial assignments permeant by adding them to flow tables of
switches. The statement \textit{Send(x)} sends specific packets to
be treated in a certain way at certain switches. To keep a record of
actions takes on packets on different switches we assume a map
called \textit{history} from the set of switches IDs to the set of
lists of pairs of packets and taken actions. This map is used in the
Rule $(\hbox{Send}^s)$. Operational semantics of these statements
are given in Figure~\ref{f3}. Judgement of inference rules in this
figure have the form
$S:(\sigma,\gamma,ir)\rightarrow(\sigma^\prime,\gamma^\prime,ir^\prime)$.
This judgement reads as following. If the execution of $S$ in the
state $(\sigma,\gamma,ir)$ ends then the execution reaches the state
$(\sigma^\prime,\gamma^\prime,ir^\prime)$.

Inference rules in Figure~\ref{f3} use that in Figure~\ref{f2} to
get the semantics of the other important construct of \textit{ImNet}
which is event transformers (et). Judgements of Figure~\ref{f2} have
the form $et:\gamma\rightarrow u$ meaning that the semantics of the
transformer $et$ in the variable state $\gamma$ is $u$. The event
transformer  $\hbox{Lift} (x,\lambda t.f(t))$ applies the map
$\lambda t.f(t)$ to values of the event in $x$ (Rule
$(\hbox{Lift}^s)$). The event transformer
$\hbox{Filter}(x,\lambda.f(t))$ filters the event in $x$ using the
map $\lambda t.f(t)$ (Rule $(\hbox{Filter}^s)$). From a given set of
actions $A$ and two events $x_1$ and $x_2$ the event transformers
$\hbox{MixFst}(A,x_1,x_2)$ and $\hbox{MixSnd}(A,x_1,x_2)$ create
lists of rules (Rules $(\hbox{Mix}_1^s)$ and $(\hbox{Mix}_2^s)$).

\begin{figure*}[t]
\centering \fbox{
\begin{minipage}{13cm}
{\footnotesize{
\[
\begin{prooftree}
et:\gamma\rightarrow u \justifies x:=
{et}:(\sigma,\gamma,ir)\rightarrow(\sigma,\gamma[x\mapsto u],ir)
\thickness=0.08em\using{(\hbox{Assgn}^s)}
\end{prooftree}
\]
\[
\begin{prooftree}
S_1:(\sigma,\gamma,ir)\rightarrow
(\sigma^{\prime\prime},\gamma^{\prime\prime},ir^{\prime\prime})\qquad
S_2: (\sigma^{\prime\prime},\gamma^{\prime\prime},ir^{\prime\prime})
\rightarrow(\sigma^\prime,\gamma^\prime,ir^\prime) \justifies S_1;
S_2:(\sigma,\gamma,ir)\rightarrow(\sigma^\prime,\gamma^\prime,ir^\prime)
\thickness=0.08em\using{(\hbox{seq}^s)}
\end{prooftree}
\]
\[
\begin{prooftree}
\gamma(x)\in \hbox{Intial-rule-assignment}\justifies
\hbox{AddRules}(x):(\sigma,\gamma,ir)\rightarrow(\sigma,\gamma,ir\cup
\gamma(x) ) \thickness=0.08em\using{(\hbox{Addrl}^s)}
\end{prooftree}
\]
\[
\begin{prooftree}
\justifies \hbox{Register}:(\sigma,\gamma,ir)\rightarrow(\sigma\cup
ir,\gamma,\emptyset) \thickness=0.08em\using{(\hbox{Reg}^s)}
\end{prooftree}
\]
\[
\begin{prooftree}
\gamma(x)=((v^1_1,v^2_1,v^3_1),(v^1_2,v^2_2,v^3_2),\ldots,(v^1_n,v^2_n,v^3_n))
\qquad \forall i. (v^i_2,v^i_3)\in \hbox{history}(v^i_1) \justifies
\hbox{Send}(x):(\sigma,\gamma,ir)\rightarrow(\sigma,\gamma,ir)
\thickness=0.08em\using{(\hbox{Send}^s)}
\end{prooftree}
\]}}
 \caption{Operational semantics statements of
ImNet}\label{f3}
\end{minipage}
}
\end{figure*}

\section{Controller Programs}
\label{s2}\vspace{-4pt}

This section presents two examples of programs constructed using the
syntax of \textit{ImNet} (Figure~\ref{f1}). The first example
constructs rules based on information stored in the variable $x$ and
then installs the established rules to flow tables of switches
stored in $z$. This program has the following statements.
\[y = \hbox{MakeRule}(x);\]
\[ z=\hbox{Lift} (z,\lambda t.(t,y));\]
\[\hbox{AddRules}(z);\]
\[Register;\]

The first statement of the program makes a rule for each value of
the event stored in $x$. Then the second statement assigns these
rules to switch IDs in the event stored in $z$. The third statement
stores the rule assignment of $z$ in $ir$ as an initial rule
assignment. The last statement of the program adds the established
rules to the flow tables of switches. Figure~\ref{f4} shows the
operational semantics of this program using the semantics of the
previous section.

\begin{figure*}
\centering \fbox{
\begin{minipage}{13cm}
{\footnotesize{
$(\emptyset,\{z\mapsto\{id_1,id_2\},x\mapsto \{((srcport(80),sendall,\_),(inport(1),sendcontroller,\_))\},[])$\\
\textbf{$y=\hbox{MakeRule}(x);$}\\
$(\emptyset,\{z\mapsto\{id_1,id_2\},x\mapsto
\{((srcport(80),sendall,\_),(inport(1),sendcontroller,\_))\},
\\y\mapsto\{(srcport(80),[sendall]),(inport(1),[sendcontroller])\}\},\emptyset)$\\
\textbf{$z=\hbox{Lift} (z,\lambda t.(t,y));$}\\
$(\emptyset,\{z\mapsto\{(id_1,\gamma(y)),(id_2,\gamma(y))\},$\\$x\mapsto
\{((srcport(80),sendall,\_),(inport(1),sendcontroller,\_))\},
\\y\mapsto\{(srcport(80),[sendall]),(inport(1),[sendcontroller])\}\}, \emptyset)$\\
\textbf{$\hbox{AddRules}(z);$}\\
$(\emptyset,\{z\mapsto\{(id_1,\gamma(y)),(id_2,\gamma(y))\},$\\$x\mapsto
\{((srcport(80),sendall,\_),(inport(1),sendcontroller,\_))\},
\\y\mapsto\{(srcport(80),[sendall]),(inport(1),[sendcontroller])\}\},
\{(id_1,\gamma(y)),(id_2,\gamma(y))\})$\\
\textbf{Register;}\\
$(\{(id_1,\gamma(y)),(id_2,\gamma(y))\},\{z\mapsto\{(id_1,\gamma(y)),(id_2,\gamma(y))\},$\\$x\mapsto
\{((srcport(80),sendall,\_),(inport(1),sendcontroller,\_))\},
\\y\mapsto\{(srcport(80),[sendall]),(inport(1),[sendcontroller])\}\},
\emptyset)$\\}} \caption{Example 1; an operational semantics of a
program written in ImNet} \label{f4}
\end{minipage}
}
\end{figure*}

The second example constructs forwarding rules based on source IPs
of arriving packets and then installs the established rules to flow
tables of switch IDs stored in $z$. This program has the following
statements.
\[y=\hbox{SourceIps};\]
\[ y=\hbox{ApplyLft} (y,\lambda t.(t,\hbox{port}(t)));\]
\[y=\hbox{Lift} (y,\lambda t.(t,\hbox{switch}(t,z));\]
\[y=\hbox{MakForwRule}(y);\]
\[\hbox{AddRules}(y);\]
\[Register;\]

The first statement of the program assumes a function
\textit{SourceIps} that returns source IPs of arriving packets and
stores them in the form of an event in $y$. The second statement
transfers event of $y$ into event of pairs of IPs and port numbers
through which packets will be forwarded. The third statement
augments values of event in $y$ with switch IDs from the event
stored in $z$. The fourth statement makes a forward rule for each
value of the event stored in $y$. Then the fifth statement stores
the rule assignment of $y$ in $ir$ as an initial rule assignment.
The last statement of the program adds the established rules to the
flow tables of switches. Figure~\ref{f5} shows the operational
semantics of this program using the semantics of the previous
section.

\begin{figure*}
\centering \fbox{
\begin{minipage}{13cm}
{\footnotesize{ $(\emptyset,\{z\mapsto\{id_1,id_2\}
\},[])$\\
\textbf{$y=\hbox{SourceIps};$}\\
$(\emptyset,\{z\mapsto\{id_1,id_2\},
\\y\mapsto\{(ip_1,pk_1),(ip_2,pk_2)\}\},\emptyset)$\\
\textbf{$y=\hbox{ApplyLft} (y,\lambda t.(t,\hbox{port}(t)));$}\\
$(\emptyset,\{z\mapsto\{id_1,id_2\},$\\
$y\mapsto\{(pr_1,pk_1),(pr_2,pk_2)\}\}, \emptyset)$\\
\textbf{$y=\hbox{Lift} (y,\lambda t.(t,\hbox{switch}(t,z));$}\\
$(\emptyset,\{z\mapsto\{id_1,id_2\},$
\\$y\mapsto\{(id_1,pr_1,pk_1),(id_2,pr_2,pk_2)\}\}, \emptyset)$\\
\textbf{$y=\hbox{MakForwRule}(y);$}\\
$(\emptyset,\{z\mapsto\{id_1,id_2\},
\\y\mapsto\{(id_1,(pk_1,sendout(pr_1)),(id_2,(pk_2,sendout(pr_2)))\}\}, \emptyset)$\\
\textbf{$\hbox{AddRules}(y);$}\\
$(\emptyset,\{z\mapsto\{id_1,id_2\},
\\y\mapsto\{(id_1,(pk_1,sendout(pr_1)),(id_2,(pk_2,sendout(pr_2)))\}\},
\\ \{(id_1,(pk_1,sendout(pr_1)),(id_2,(pk_2,sendout(pr_2)))\})$\\
\textbf{Register;}\\
$(\{(id_1,(pk_1,sendout(pr_1)),(id_2,(pk_2,sendout(pr_2)))\},\{z\mapsto\{id_1,id_2\},
\\y\mapsto\{(id_1,(pk_1,sendout(pr_1)),(id_2,(pk_2,sendout(pr_2)))\}\}, \emptyset)$}}
\caption{Example 2; an operational semantics of a program written in
ImNet} \label{f5}
\end{minipage}
}
\end{figure*}

\section{Related and Future Work}
\label{s3}\vspace{-4pt}

This section presents work most related to that presented in the
current paper.

One of the early attempts to develop software-defined networking
(SDN) is NOX~\cite{Gude08} based on ideas from~\cite{Casado09} and
4D~\cite{Greenberg05}. On the switch-level, NOX uses explicit and
callbacks rules for packet-processing. Examples of applications that
benefitted from NOX are load balancer~\cite{Wang11} and the work
in~\cite{Handigol09,Heller10}. Many directions for improving
platforms of programming networks include Maestro~\cite{Cai10} and
Onix~\cite{Koponen10}, which uses distribution and parallelization
to provide better performance and scalability.

A famous programming language for networks is
Frenetic~\cite{Foster10,Foster11} which has two main components. The
first component is a collection of operators that are source-level.
The operators aim at establishing and treating streams of network
traffic. These operators also are built on concepts of functional
programming (FP) and query languages of declarative database.
Moreover the operators support a modular design, a cost control, a
race-free semantics, a single-tier programming, and a declarative
design. The second component of Frenetic is a run-time system. This
system facilitates all of the actions of adding and removing
low-level rules to and from flow tables of switches. One advantage
of {\em ImNet}, the language presented in this paper, over Frenetic
is that {\em ImNet} is imperative. Therefore  {\em ImNet}  paves the
way to the appearance of other types of network programming
languages such as object-oriented network programming langues and
context-oriented network programming languages.

Other examples to program network components though high-level
languages are NDLog and NetCore~\cite{Loo05netcore}. NetCore
provides an integrated view of the whole network. NDLog is designed
in an explicitly distributed fashion.

As an extension of Datalog, NDLog~\cite{Loo05,Loo05-2} was presented
to determine and code protocols of routing~\cite{Arneson}, overlay
networks, and concepts like hash tables of  distributed systems.
{\em ImNet} (presented in this paper), Frenetic, and NDLog can be
classified as high-level network programming languages. While NDLog
main focus is overlay networks and routing protocols, Frenetic (in a
functional way) and {\em ImNet} (in an imperative way) focus on
implementing packet processing such as modifying header fields.
Therefore {\em ImNet} equips a network programmer with a modular
view of the network which is not provided by NDLog and Frenetic.
This is supported by the fact that  a program in NDLog is a single
query that is calculated on each router of the network.

The switch component~\cite{McKeown} of networks can be programmed
via many interfaces such as OpenFlow platform. Examples of other
platforms include Shangri-La~\cite{Chen} and FPL-3E~\cite{Cristea},
RouteBricks~\cite{Dobrescu}, Click modular router~\cite{Kohler},
Snortran~\cite{Egorov} and Bro~\cite{Paxson}. The idea in
Shangri-La~\cite{Chen} and FPL-3E~\cite{Cristea} is to produce
certain hardware for packet-processing from high-level programs that
achieves packet-processing. In RouteBricks~\cite{Dobrescu}, stock
machines are used to improve performance of program switches. As a
modular approach, the platform of Click modular
router~\cite{Kohler}, enables programming network components.  This
system focuses on software switches in the form of Linux kernel
code. For the sake of intrusions detection and preserving network
security, Snortran~\cite{Egorov} and Bro~\cite{Paxson} enable coding
monitoring strategies and robust packet-filtering. One advantage of
{\em ImNet}, the language presented in this paper, over all the
related work is that {\em ImNet} overcomes the disadvantage of most
similar languages of focusing on controlling a single device.

There are many interning directions for future work. One such
direction is develop methods for static analysis of network
programming languages. Obviously associating these analyses with
correctness proofs, in the spirit
of~\cite{El-Zawawy11-2,El-Zawawy11-4,El-Zawawy13-1,El-Zawawy13-2},
will have many network applications.

\section{Conclusion}
\label{s4}\vspace{-4pt}

Software-Defined Networks (SDNs) is a recent architectures of
networks in which a controller device programs other network devices
(specially switches) via a sequence of installing and uninstalling
rules to memories of these devices.

In this paper, we presented a high-level imperative network
programming language, called \textit{ImNet}, to facilitate the job
of controller through efficient, yet simple, programs.
\textit{ImNet} has the advantages of simplicity, expressivity,
propositionally, and being imperative. The paper also introduced a
concrete operational semantics to meanings of \textit{ImNet}
constructs. Detailed examples of using \textit{ImNet} and the
operational semantics were also illustrated in the paper.

\end{document}

%% file: prooftree.tex
\newdimen\proofrulebreadth \proofrulebreadth=.05em
\newdimen\proofdotseparation \proofdotseparation=1.25ex
\newdimen\proofrulebaseline \proofrulebaseline=2ex
\newcount\proofdotnumber \proofdotnumber=3
\let\then\relax
\def\hfi{\hskip0pt plus.0001fil}
\mathchardef\squigto="3A3B
\newif\ifinsideprooftree\insideprooftreefalse
\newif\ifonleftofproofrule\onleftofproofrulefalse
\newif\ifproofdots\proofdotsfalse
\newif\ifdoubleproof\doubleprooffalse
\let\wereinproofbit\relax
\newdimen\shortenproofleft
\newdimen\shortenproofright
\newdimen\proofbelowshift
\newbox\proofabove
\newbox\proofbelow
\newbox\proofrulename
\def\shiftproofbelow{\let\next\relax\afterassignment\setshiftproofbelow\dimen0 }
\def\shiftproofbelowneg{\def\next{\multiply\dimen0 by-1 }%
\afterassignment\setshiftproofbelow\dimen0 }
\def\setshiftproofbelow{\next\proofbelowshift=\dimen0 }
\def\setproofrulebreadth{\proofrulebreadth}

\def\prooftree{
\ifnum  \lastpenalty=1
\then   \unpenalty
\else   \onleftofproofrulefalse
\fi
\ifonleftofproofrule
\else   \ifinsideprooftree
        \then   \hskip.5em plus1fil
        \fi
\fi
\bgroup
\setbox\proofbelow=\hbox{}\setbox\proofrulename=\hbox{}%
\let\justifies\proofover\let\leadsto\proofoverdots\let\Justifies\proofoverdbl
\let\using\proofusing\let\[\prooftree
\ifinsideprooftree\let\]\endprooftree\fi
\proofdotsfalse\doubleprooffalse
\let\thickness\setproofrulebreadth
\let\shiftright\shiftproofbelow \let\shift\shiftproofbelow
\let\shiftleft\shiftproofbelowneg
\let\ifwasinsideprooftree\ifinsideprooftree
\insideprooftreetrue
\setbox\proofabove=\hbox\bgroup$\displaystyle 
\let\wereinproofbit\prooftree
\shortenproofleft=0pt \shortenproofright=0pt \proofbelowshift=0pt
\onleftofproofruletrue\penalty1
}

\def\eproofbit{
\ifx    \wereinproofbit\prooftree
\then   \ifcase \lastpenalty
        \then   \shortenproofright=0pt  
        \or     \unpenalty\hfil         
        \or     \unpenalty\unskip       
        \else   \shortenproofright=0pt  
        \fi
\fi
\global\dimen0=\shortenproofleft
\global\dimen1=\shortenproofright
\global\dimen2=\proofrulebreadth
\global\dimen3=\proofbelowshift
\global\dimen4=\proofdotseparation
\global\count255=\proofdotnumber
$\egroup  
\shortenproofleft=\dimen0
\shortenproofright=\dimen1
\proofrulebreadth=\dimen2
\proofbelowshift=\dimen3
\proofdotseparation=\dimen4
\proofdotnumber=\count255
}

\def\proofover{
\eproofbit 
\setbox\proofbelow=\hbox\bgroup 
\let\wereinproofbit\proofover
$\displaystyle
}%
\def\proofoverdbl{
\eproofbit 
\doubleprooftrue
\setbox\proofbelow=\hbox\bgroup 
\let\wereinproofbit\proofoverdbl
$\displaystyle
}%
\def\proofoverdots{
\eproofbit 
\proofdotstrue
\setbox\proofbelow=\hbox\bgroup 
\let\wereinproofbit\proofoverdots
$\displaystyle
}%
\def\proofusing{
\eproofbit 
\setbox\proofrulename=\hbox\bgroup 
\let\wereinproofbit\proofusing
\kern0.3em$
}

\def\endprooftree{
\eproofbit 
  \dimen5 =0pt
\dimen0=\wd\proofabove \advance\dimen0-\shortenproofleft
\advance\dimen0-\shortenproofright
\dimen1=.5\dimen0 \advance\dimen1-.5\wd\proofbelow
\dimen4=\dimen1
\advance\dimen1\proofbelowshift \advance\dimen4-\proofbelowshift
\ifdim  \dimen1<0pt
\then   \advance\shortenproofleft\dimen1
        \advance\dimen0-\dimen1
        \dimen1=0pt
        \ifdim  \shortenproofleft<0pt
        \then   \setbox\proofabove=\hbox{%
                        \kern-\shortenproofleft\unhbox\proofabove}%
                \shortenproofleft=0pt
        \fi
\fi
\ifdim  \dimen4<0pt
\then   \advance\shortenproofright\dimen4
        \advance\dimen0-\dimen4
        \dimen4=0pt
\fi
\ifdim  \shortenproofright<\wd\proofrulename
\then   \shortenproofright=\wd\proofrulename
\fi
\dimen2=\shortenproofleft \advance\dimen2 by\dimen1
\dimen3=\shortenproofright\advance\dimen3 by\dimen4
\ifproofdots
\then
        \dimen6=\shortenproofleft \advance\dimen6 .5\dimen0
        \setbox1=\vbox to\proofdotseparation{\vss\hbox{$\cdot$}\vss}%
        \setbox0=\hbox{%
                \advance\dimen6-.5\wd1
                \kern\dimen6
                $\vcenter to\proofdotnumber\proofdotseparation
                        {\leaders\box1\vfill}$%
                \unhbox\proofrulename}%
\else   \dimen6=\fontdimen22\the\textfont2 
        \dimen7=\dimen6
        \advance\dimen6by.5\proofrulebreadth
        \advance\dimen7by-.5\proofrulebreadth
        \setbox0=\hbox{%
                \kern\shortenproofleft
                \ifdoubleproof
                \then   \hbox to\dimen0{%
                        $\mathsurround0pt\mathord=\mkern-6mu%
                        \cleaders\hbox{$\mkern-2mu=\mkern-2mu$}\hfill
                        \mkern-6mu\mathord=$}%
                \else   \vrule height\dimen6 depth-\dimen7 width\dimen0
                \fi
                \unhbox\proofrulename}%
        \ht0=\dimen6 \dp0=-\dimen7
\fi
\let\doll\relax
\ifwasinsideprooftree
\then   \let\VBOX\vbox
\else   \ifmmode\else$\let\doll=$\fi
        \let\VBOX\vcenter
\fi
\VBOX   {\baselineskip\proofrulebaseline \lineskip.2ex
        \expandafter\lineskiplimit\ifproofdots0ex\else-0.6ex\fi
        \hbox   spread\dimen5   {\hfi\unhbox\proofabove\hfi}%
        \hbox{\box0}%
        \hbox   {\kern\dimen2 \box\proofbelow}}\doll%
\global\dimen2=\dimen2
\global\dimen3=\dimen3
\egroup 
\ifonleftofproofrule
\then   \shortenproofleft=\dimen2
\fi
\shortenproofright=\dimen3
\onleftofproofrulefalse
\ifinsideprooftree
\then   \hskip.5em plus 1fil \penalty2
\fi
}
